\documentclass[
 aip,
 jap,
 amsmath,amssymb,
 reprint,
]{revtex4-1}

\draft 

\usepackage{graphicx} 
\usepackage{dcolumn} 
\usepackage{bm} 

\usepackage{hyperref}
\usepackage[utf8]{inputenc}
\usepackage[T1]{fontenc}
\usepackage{mathptmx}
\usepackage{etoolbox}
\makeatletter
\def\@email#1#2{%
 \endgroup
 \patchcmd{\titleblock@produce}
  {\frontmatter@RRAPformat}
  {\frontmatter@RRAPformat{\produce@RRAP{*#1\href{mailto:#2}{#2}}}\frontmatter@RRAPformat}
  {}{}
}%
\makeatother
\begin{document}

\title[Electrical Noise Produced by Micron-Sized Particles above a Surface Paul Trap]{Electrical Noise Produced by Micron-Sized Particles above a Surface Paul Trap} 

\author{Ben Saarel}
\email[]{bsaarel@gmail.com}
\affiliation{Department of Physics, University of California, Berkeley, CA 94720, USA}
\affiliation{Challenge Institute for Quantum Computation, University of California, Berkeley, CA 94720, USA}

\author{Ozgur Sahin}
\affiliation{Department of Chemistry, University of California, Berkeley, CA 94720, USA}
\affiliation{Challenge Institute for Quantum Computation, University of California, Berkeley, CA 94720, USA}

\author{Hartmut H\"affner}
\email[]{hhaeffner@berkeley.edu}
\affiliation{Department of Physics, University of California, Berkeley, CA 94720, USA}
\affiliation{Challenge Institute for Quantum Computation, University of California, Berkeley, CA 94720, USA}

\author{Alpha T. N'Diaye}
\affiliation{Advanced Light Source, Lawrence Berkeley National Laboratory, Berkeley, CA 94720, USA}

\date{\today}

\begin{abstract}
Electric field noise produced by the surface of ion trap electrodes reduces the fidelity of quantum computing operations. Despite decades of investigation its microscopic origins remain unclear. Here, we measure electric field noise at trapping locations along the symmetry axis of a linear surface Paul trap. We find that noise levels vary by three orders-of-magnitude in one 600$\,$\textmu m section of the trap. Optical and scanning electron microscope images show micron-sized particles close to the trapping locations with the highest noise levels. We find that modeling the particles as a lossy dielectric with a effective loss tangent $\tan\theta=0.33(0.06)$ describes the magnitude of the noise, as well as its spatial and frequency dependence. Our observations may explain the large variation of reported noise levels in literature. 
\end{abstract}

\maketitle 

\section{Introduction}
Ion trap quantum computers use conductive electrodes to control the position and motion of ions. Computing operations that depend on ion motion can be sped up by bringing the ions close to these electrodes \cite{brown_materials_2021}. However, as the ions are brought closer to the electrodes, they experience electric field noise with a magnitude that is much greater than predicted by known noise mechanisms \cite{turchette_heating_2000,brownnutt_ion-trap_2015}. Importantly, spectral components of the noise near the motional frequency of the ions cause motional heating, and this heating acts as a source of decoherence that lowers the fidelity of quantum operations \cite{ballance_high-fidelity_2016,harty_high-fidelity_2016}. Determining the origins of the noise will make it easier to reliably identify and mitigate the noise so faster, higher fidelity operations can be performed. There is also a fundamental scientific interest in identifying the unknown noise mechanisms to gain a better understanding of the physics that takes place at the surfaces of materials. 

In this paper, we measure electric field noise along a 1.2$\,$mm segment of the symmetry axis of a linear surface Paul trap. In one 600$\,$\textmu m section of the trap, the noise amplitude varies by three orders-of-magnitude. Surface characterization reveals micron-scale particles near the trapping sites with the highest noise levels. By modeling these particles as a lossy dielectric with an effective loss tangent of $\tan\theta=0.33(0.06)$, we can account for the observed noise magnitude as well as its spatial and frequency dependence. 

\section{Methods}
\subsection{Trap Fabrication}
The linear surface Paul trap we use in this study is fabricated using the following steps. Trenches 20$\,$\textmu m in width and 100$\,$\textmu m in depth are etched into a fused silica substrate to define the electrode geometry \cite{femtika_selective_laser_etching,daniilidis_surface_2014}. This trench-etching step is performed by an outside company (Femtika), after which the trap is transported to a class 100 cleanroom (UC Berkeley's Marvell Nanolab) to coat the electrodes with metal. 

\begin{figure*}[ht]
\centering
\includegraphics[width=6.69in,scale=1.0]{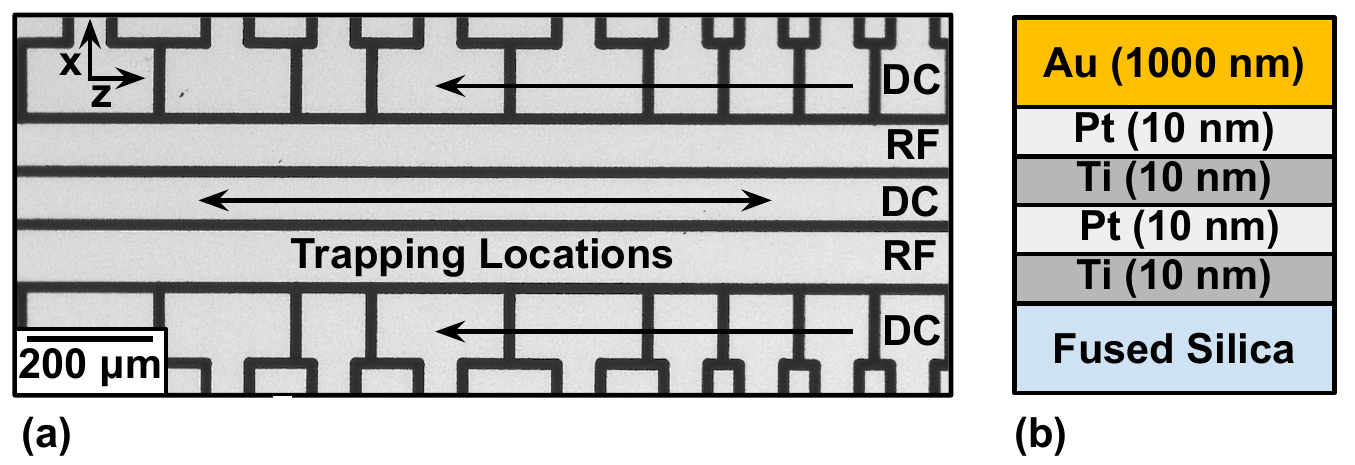}
\caption{Trap fabrication and layout. (a) Microscope image of the trap after fabrication. (b) Schematic of the metal layers deposited onto the fused silica substrate. }
\label{fig:trap-layout}
\end{figure*}

We use an electron-beam evaporator to deposit layers of 10$\,$nm titanium, 10$\,$nm platinum, and 1000$\,$nm gold to make the electrodes conductive. The metal layers are deposited at a $30^\circ$ angle with respect to the trap surface so that the bottom of the trenches remain non-conductive; this ensures that the electrodes are electrically isolated from one another. The titanium serves as a sticking layer that allows the other metals to adhere to the surface. The platinum layer acts as a diffusion barrier to mitigate titanium diffusion into the gold layers when the trap is heated. Fig.\,\ref{fig:trap-layout}a. displays a microscope image of the trap after fabrication. Fig.\,\ref{fig:trap-layout}b shows a schematic cross section of the electrodes with the various metal layers that make up the trap. 

After fabrication, we store the trap in a vacuum desiccator for four months. To measure the noise above the trap surface, we remove the trap from the desiccator, wirebond it in the cleanroom, and install it into the experiment vacuum chamber we use for noise measurements. To mitigate dust accumulation on the trap in the lab atmosphere, we construct a plastic tent around our vacuum chamber and filter the air within the tent using an air purifier with fresh HEPA filters. We estimate the total time in the cleanroom to be a few hours and the total time in the lab atmosphere to be $<20$ minutes. To further minimize contamination, we wear clean-room suites and face masks throughout the fabrication and installation process. Prior to noise measurements we heat the trap in the experiment vacuum chamber to 450$\,$C for one hour and to 150$\,$C for four days \footnote{This high-temperature heating step recrystallizes the trap surface. The results of this recrystallization will be discussed in a future publication.}. 

\subsection{Noise Measurements}
To probe the noise above the surface of the trap, we measure the heating rate of a single $^{40}$Ca$^+$ ion's $z$-mode of motion. The heating rate of the ion's motion $\dot{\overline{n}}$ is directly proportional to the spectral density of noise $S_\mathrm{E}(\omega)$ \cite{brownnutt_ion-trap_2015}:
\begin{equation}
S_E(\omega) = \dot{\overline{n}}(\omega)\times\frac{4 m \hbar \omega }{q^2},
\end{equation}
where $q$ is the charge of the ion, $m$ is the ion mass, and $\omega$ is the frequency of the motional mode. 

To measure the heating rate of the ion's $z$-motion, we first Doppler cool the motion using 397$\,$nm laser light that is red-detuned from the $\mathrm{S}_{1/2} \leftrightarrow \mathrm{P}_{1/2}$ transition of $^{40}$Ca$^+$. We then switch off the cooling light and measure the increase in the mean $z$-motional occupation number, $\overline{n}$, as a function of time. 

We determine $\overline{n}$ by driving resonant Rabi oscillations on either the \{S$_{1/2},m_j=-1/2\}
\leftrightarrow$
\{D$_{5/2},m_j=-5/2$\} or \{S$_{1/2},m_j=-1/2\}
\leftrightarrow$
\{D$_{5/2},m_j=-3/2$\} transition using 729$\,$nm laser light. We align the 729$\,$nm laser to be nearly parallel to the $z$-axis so that the Rabi oscillations are sensitive primarily to the $z$-motional mode; we estimate the angle between the laser propagation direction and the $z$-mode to be $\approx1^\circ$ for our heating rate measurements. The D-state population $P_\mathrm{D}(t)$ depends on the number of $z$-motional excitations $n$ according to  \cite{leibfried2003quantum}:
\begin{eqnarray}\label{eqn:Rabi}
P_\mathrm{D}(t) &=& \sum_n P_n \sin^2(\Omega_n t),\nonumber\\
\Omega_n &=& \ \Omega_0e^{-\eta^2/2}L_n(\eta^2)
\end{eqnarray}
where $\Omega_n$ is the $n$-dependent Rabi frequency, $\eta$ is the Lamb Dicke parameter, $L_n(\eta^2)$ is the Laguerre polynomial of degree $n$\cite{leibfried2003quantum}, and $P_n$ is the probability of measuring $n$ motional excitations which we assume to follow a thermal distribution \cite{leibfried2003quantum}:
\begin{equation}\label{eqn:Pn}
P_n = \frac{1}{\overline{n}+1}\left(\frac{\overline{n}}{\overline{n}+1}\right)^n. 
\end{equation}
We determine $\overline{n}$ by fitting the Rabi oscillations using Eqs.\,\ref{eqn:Rabi} and \ref{eqn:Pn}. 

To extract the frequency dependence of the noise, we measure heating rates at motional frequencies ranging from 0.6-1.4\,MHz and fit a power law to the heating rates of the form:
\begin{equation}\label{eq:frequency-scaling}
\dot{\overline{n}}(\omega) = \dot{\overline{n}}(\omega_0)\times\bigg(\frac{\omega_0}{\omega}\bigg)^{\alpha+1},
\end{equation}
where $\dot{\overline{n}}(\omega_0)$ is the heating rate at motional frequency $\omega_0=2\pi\times1\,$MHz, and $\alpha$ is the frequency scaling exponent of the noise ($S_\mathrm{E}\propto 1/\omega^\alpha$). When fitting the data to Eq.~\ref{eq:frequency-scaling} we vary $\dot{\overline{n}}(\omega_0)$ and $\alpha$ to obtain the best fit. 

\section{Results}
\subsection{Heating Rate Measurements}
\begin{figure}[ht]
\centering
\includegraphics[width=3.37in,scale=1.0]{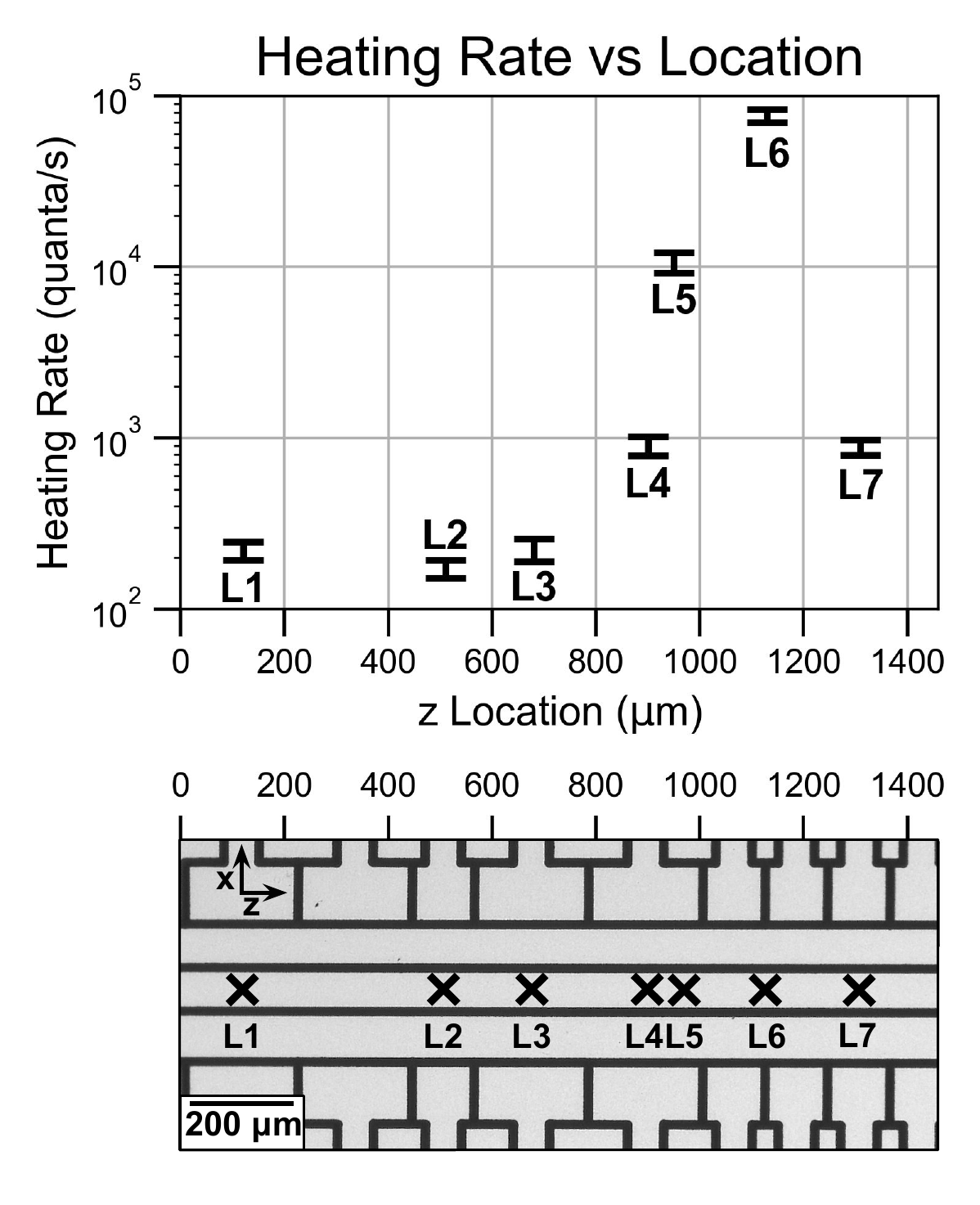}
\caption{Heating rates measured at different $z$-locations along the trap. All heating rates are taken at a motional frequency $\omega=2\pi\times  1\,$MHz. }
\label{fig:hr_vs_location}
\end{figure}

Fig.\,\ref{fig:hr_vs_location} displays the measured heating rates for $\omega=2\pi \times 1$\,MHz at various trapping locations along the length of the trap.  For positions L1-L3, the heating rates remain constant at  $\approx200\,$quanta/s. For positions L3-L7, the heating rate depends strongly on the trapping location $z$. The maximum heating rate we measure is 77000(7000) quanta/s at location L6. 

\begin{figure}[ht]
    \centering
    \includegraphics[width=3.37in,scale=1.0]{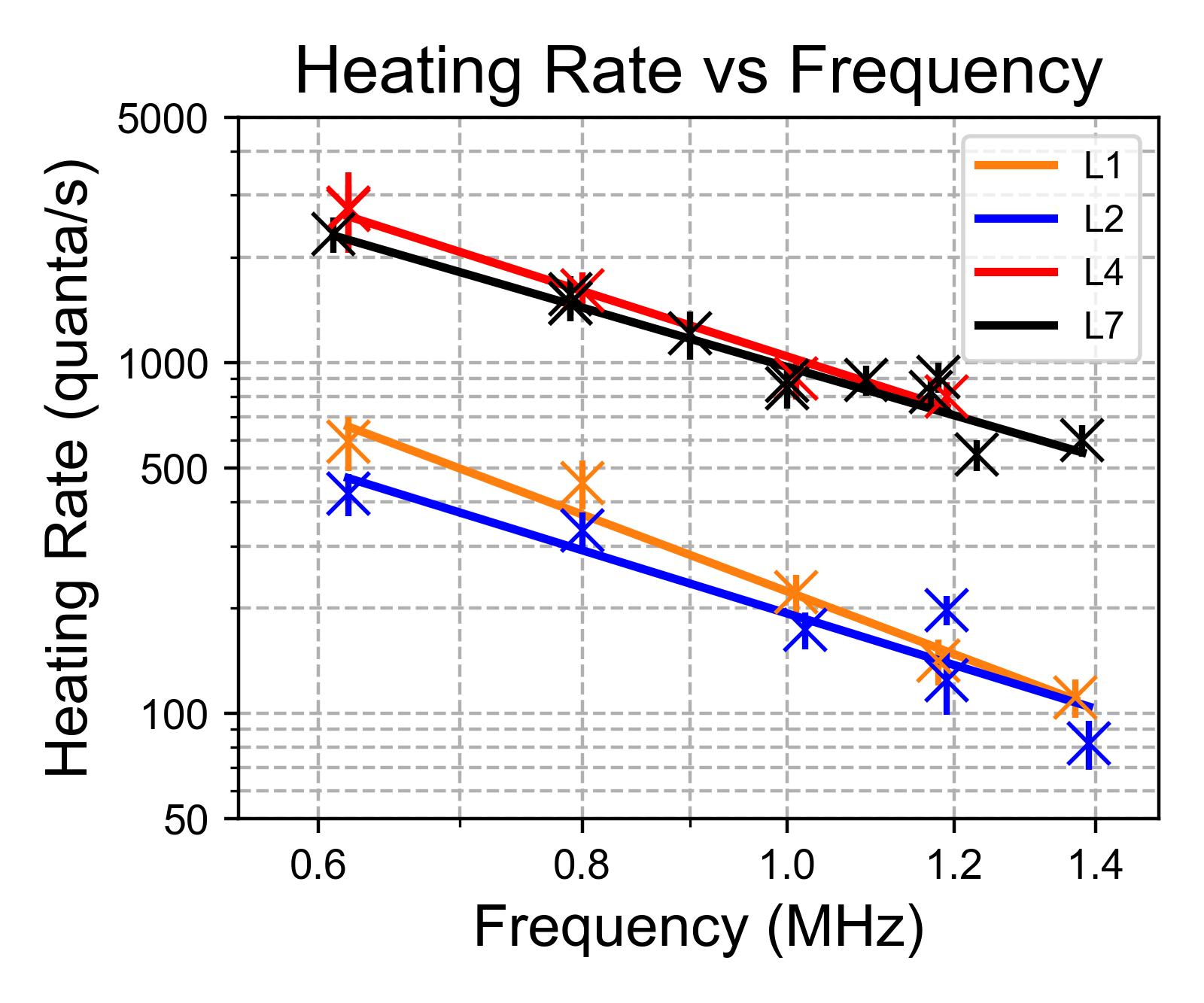}
    \caption{Frequency dependence of the heating rates at trapping locations L1, L2, L4, and L7. We fit Eq.\,\ref{eq:frequency-scaling} to the data and plot the best-fit curves in the figure. }
    \label{fig:hr_vs_freq}
\end{figure}

Fig.\,\ref{fig:hr_vs_freq} shows the scaling of the heating rate with motional frequency $\omega$ at selected locations along the trap. We fit Eq.\,\ref{eq:frequency-scaling} to determine the power-law noise exponent $\alpha$; the best-fit curves are plotted in Fig.\,\ref{fig:hr_vs_freq} and the fit parameters are displayed in Table~\ref{tab:freq}. 

\begin{table}[ht]
\caption{\label{tab:freq}
Fit parameters obtained by fitting Eq.\,\ref{eq:frequency-scaling} to the data in Fig.\,\ref{fig:hr_vs_freq}.}
\begin{ruledtabular}
\begin{tabular}{c c c}
Location & $\dot{\overline{\mathrm{n}}}(\omega_0)$ (quanta/s) & $\alpha$ \\ [0.5ex] \hline & \\[-2.0ex]
L1  & 220(10) & 1.3(2) \\ [0.5ex] 
L2  & 190(10) & 0.9(2) \\ [0.5ex] 
L4  & 1040(70) & 0.9(3) \\ [0.5ex] 
L7  & 970(30) & 0.7(1) \\ [0.5ex] 

\end{tabular}
\end{ruledtabular}
\end{table}

\subsection{Trap Characterization}
\begin{figure*}[ht]
\centering
\includegraphics[width=6.69in,scale=1.0]{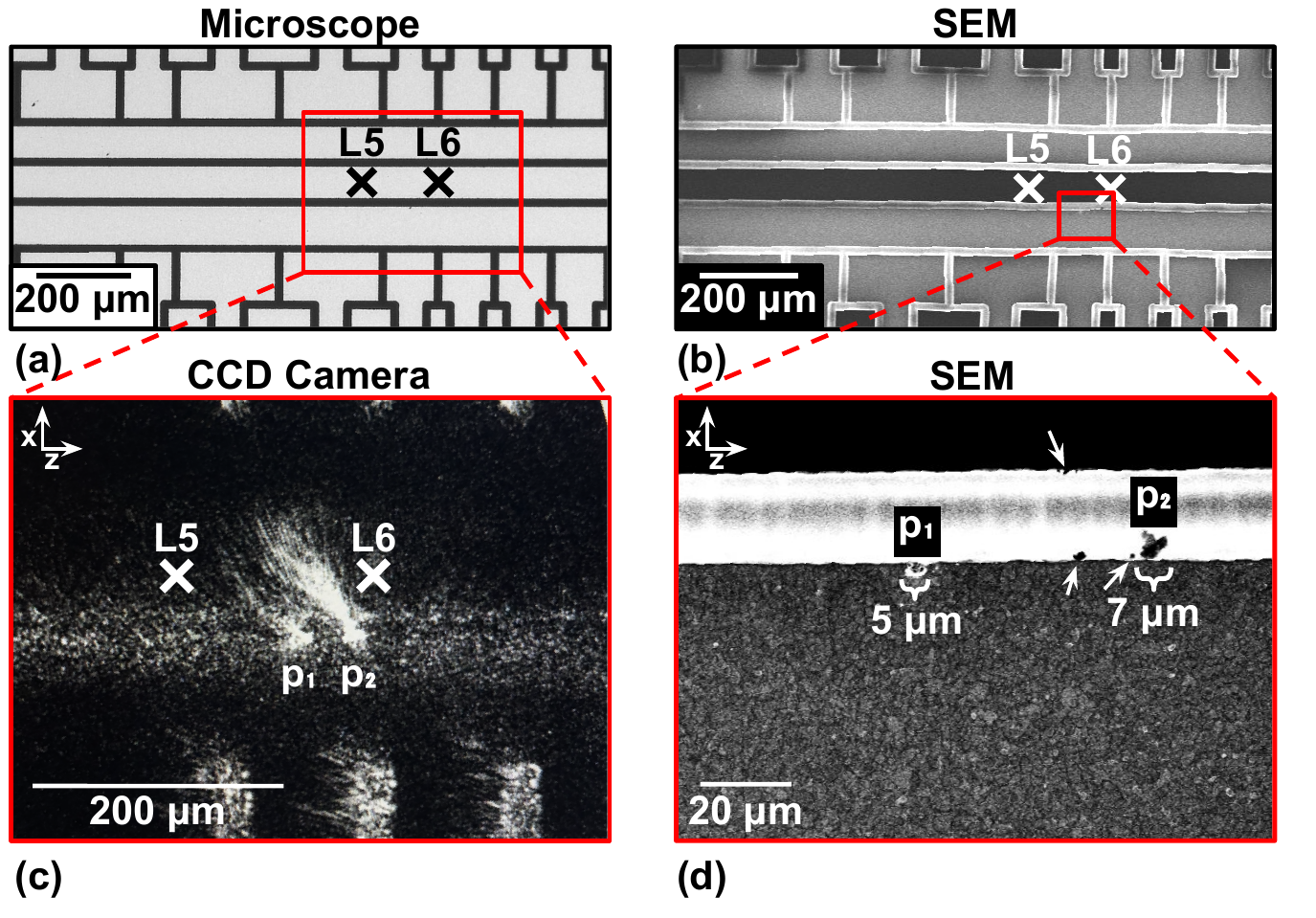}
\caption{Defect images. (a) Microscope image of the trap immediately after fabrication and prior to heating rate measurements. (b) Low-magnification SEM image of the trap taken after heating rate measurements. (c) CCD camera image of the trap taken during heating rate measurements, with point defects p$_1$ and p$_2$ illuminated by a grazing-incidence laser beam. (d) High-magnification SEM image of the trap taken after heating rate measurements. The two largest particles are labeled p$_1$ and p$_2$ and correspond to the point defects in the CCD camera image. Smaller particles are indicated by arrows. }
\label{fig:defect}
\end{figure*}

During heating rate measurements, we observe two point defects scattering laser light in the region with elevated noise levels (L4--L7). Fig.~\ref{fig:defect}c shows a CCD camera image of these defects scattering laser light inside the vacuum chamber, with the two point defects labeled p$_1$ and p$_2$. 

To image the defects with higher resolution, we remove the trap from the experimental vacuum chamber and examine it using a scanning electron microscope (SEM). Fig.~\ref{fig:defect}d shows an SEM image of the region containing point defects p$_1$ and p$_2$, revealing that they are particles several micrometers in diameter and located along the edge of a trap electrode. We also detect smaller particles in the SEM images, which are indicated by arrows. 

Particles p$_1$ and p$_2$ are located at positions $z(\mathrm{p_1})=1069\,$\textmu m and $z(\mathrm{p_2})=1114\,$\textmu m, which places them between trapping locations L5 and L6 in Fig.\,\ref{fig:hr_vs_location}. We do not have higher-resolution images or surface composition measurements of the particles because they vanished from view after a few minutes of SEM imaging. We suspect that the SEM charged the particles and caused them to leave their initial location. 

\section{Discussion}
We conjecture that the particles seen in the CCD and SEM images are responsible for the elevated noise levels at locations L4-L7 because the noise correlates with the position of the particles. The measured frequency dependence of the noise at locations thought to be limited by the particles is $S_\mathrm{E}(\omega)\propto 1/\omega^{\alpha}$, with $\alpha = 0.9(3)$ (L4) and $\alpha = 0.7(1)$ (L7).

We also note that the particles are not the dominant noise source for locations L1-L3 because noise levels are constant across these locations. Further, the heating rates at positions L1-L3 are state-of-the-art for a room-temperature, untreated trap \cite{brownnutt_ion-trap_2015}. This indicates that this region of the trap is limited by the same noise sources that limit other low-noise traps. The measured frequency dependence of the noise at locations in this low-noise region is $S_\mathrm{E}(\omega)\propto 1/\omega^{\alpha}$, with $\alpha = 1.3(2)$ (L1) and $\alpha=0.9(2)$ (L2).

\subsection{Modelling Particles as a Lossy Dielectric}
Next, we model the particles p$_1$ and p$_2$ as small chunks of lossy dielectric material and compare the model's predictions to our heating rate measurements. We adopt this model because it predicts a frequency dependence that is close to our observations, $S_\mathrm{E}(\omega) \propto 1/\omega$, provided that the effective loss tangent of the particles $\tan\theta$ depends only weakly on the frequency.

To determine the noise generated by lossy dielectric particles, we treat the ion as a point charge that oscillates at frequency $\omega$ with a (classical) motional amplitude along the z-direction of $\delta z$. It produces an electric field $\textit{\textbf{E}}=\textit{\textbf{E}}_0+\mathrm{Re}[\textit{\textbf{E}}_1 e^{i\omega t}]$, where $\textit{\textbf{E}}_0$ is the field associated with the ion’s average position and $\mathrm{Re}[\textit{\textbf{E}}_1 e^{i\omega t}]$ is the time-dependent component of the field \cite{teller_heating_2021}. When a dielectric with permittivity $\epsilon=\epsilon_0\epsilon_r(1+i\tan\theta)$ is placed near the ion, where $\epsilon_0$ is the permittivity of vacuum, $\epsilon_r$ is the relative permittivity of the dielectric, and $\tan\theta$ is the loss tangent of the dielectric, the electrical power dissipated within the dielectric is \cite{teller_heating_2021}:
\begin{equation}
\overline{P}_\mathrm{loss} = \frac{\omega}{2}\epsilon_0\epsilon_r\tan\theta\int_V |\textbf{\textit{E}}_1|^2\,dV.
\end{equation}
$\overline{P}_\mathrm{loss}$ is the cycle-averaged power dissipated within the dielectric, the integral is evaluated over the dielectric's volume $V$, and we assume that $\delta z$ is small compared to the ion-dielectric separation. The fluctuation–dissipation theorem relates the dissipated power to the thermal noise emitted by the dielectric:
\begin{eqnarray}\label{eq:thermal_noise}
S_\mathrm{E}(\omega) &=& \frac{8 k_\mathrm{B}T}{(\delta zq \omega)^2}\times\overline{P}_\mathrm{loss} 
\nonumber\\ &=& \frac{4 k_\mathrm{B}T}{ \delta z^2 q^2 \omega}\epsilon_0\epsilon_r\tan\theta \int_V |\textbf{\textit{E}}_1|^2\,dV\,,
\end{eqnarray}
where $T=293\,$K for our room-temperature trap. 

\begin{figure*}[ht]
    \centering
    \includegraphics[width=6.69in,scale=1.0]{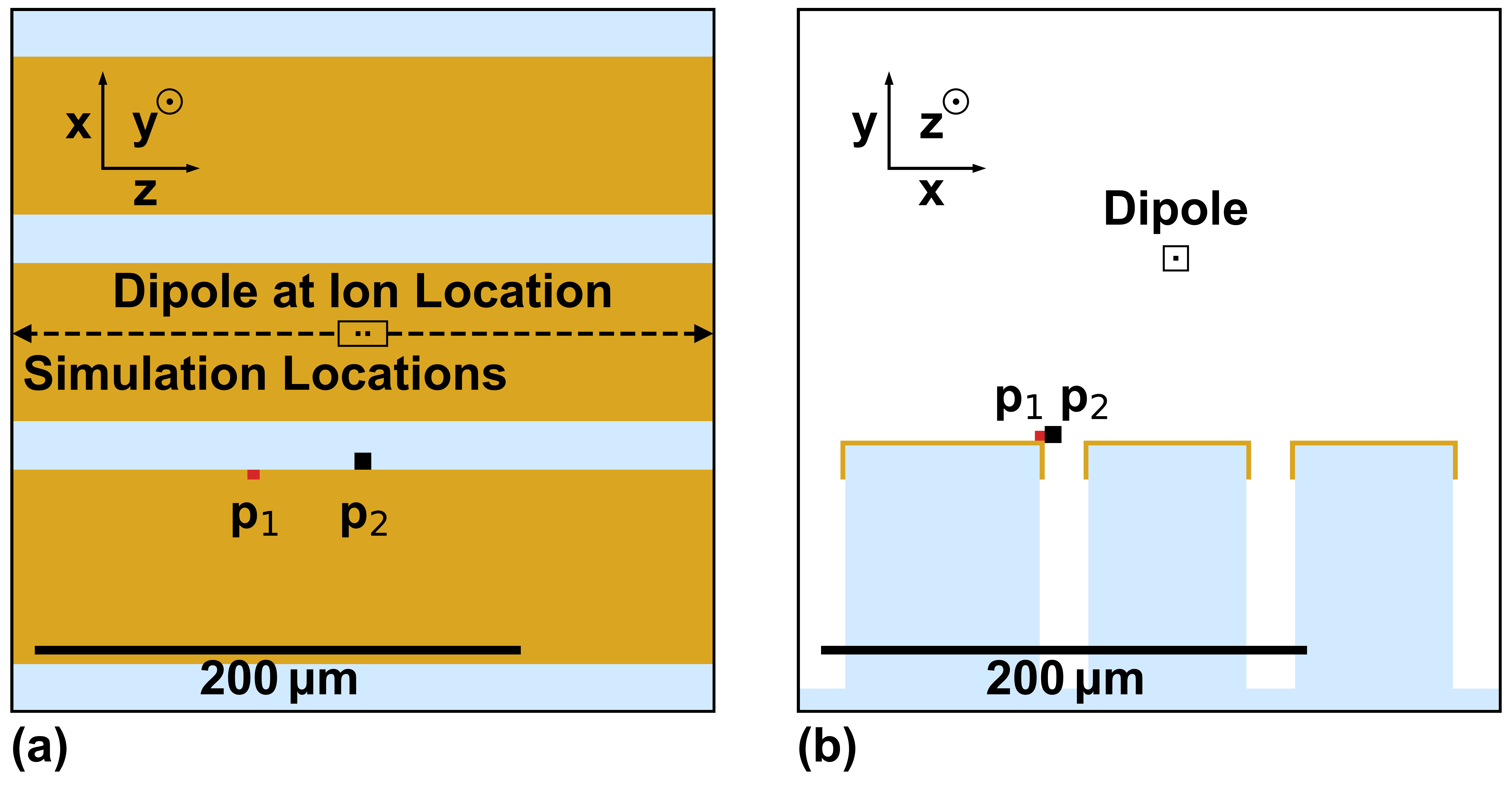}
    \caption{Simulation setup. (a) Top view of the simulation showing the dipole and model particles. (b) Side view of the simulation model. All components are drawn to scale except for the $\pm q$ charges that make up the dipole: their separation is to scale, but their size is exaggerated for visibility. In the figure, the charges are shown as $2\times2\times2\,$\textmu m cubes, whereas in the simulation they are $0.4\times0.4\times0.4\,$\textmu m.}
    \label{fig:simulation_setup}
\end{figure*}

We determine $\textit{\textbf{E}}_1$ using a finite element 
analysis (FEA) electrostatic field simulation of the trap and particles (Fig.$\,$\ref{fig:simulation_setup}). 
To calculate the electric field within the simulation ($\textit{\textbf{E}}_\mathrm{1,sim}$) we place a dipole consisting of two opposite charges $\pm q$ with separation $\delta z =4\,$\textmu m at the ion location above the trap. We model the two largest particles p$_1$ and p$_2$ as cubes with dimensions $(4 \times 4 \times 5)\,$\textmu$\text{m}^3$ and $(7 \times 7 \times 7)\,$\textmu$\text{m}^3$, respectively--- we determine the $x$- and $z$-dimensions of the particles within the simulation using Fig.\,\ref{fig:defect}d and we assume the $y$-dimension of each particle is equal to its $x$-dimension. We also use Fig.\,\ref{fig:defect}d to determine the particle positions on the trap surface. We assume p$_2$ is at the top corner of the RF electrode trench (not embedded deep in the trench) because it scatters significant amounts of grazing-incidence laser light. 

We assume that the relative permittivity of the two particles is $\epsilon_r = 5$, which is common for polymers \cite{ohki_2022,tan_2006,nasrin_effect_2020,lee_robust_2017}. We vary the $z$-location of the dipole ($z_\mathrm{I}$) along the length of the trap in the simulation to create a parametrized model of the expected heating rates as a function of $z_\mathrm{I}$: 
\begin{eqnarray}\label{eq:linear_fit}
\dot{\overline{n}}(z_\mathrm{I}) &=& \tan\theta \times \bigg(\frac{q^2}{4 m\hbar\omega}\bigg)\times\bigg(\frac{4 k_\mathrm{B}T}{ \delta z^2 q^2 \omega}\epsilon_0\epsilon_r \bigg) \nonumber\\ &&\times \left(\int_V |\textbf{\textit{E}}_{1,\mathrm{sim}}|^2 dV\right) \Bigg|_{z_\mathrm{I}} + \dot{\overline{n}}_\mathrm{b}\,,
\end{eqnarray}
where $\tan\theta$ is the effective loss tangent of the particles and $\dot{\overline{n}}_\mathrm{b}$ is the background heating rate, both of which are treated as free fit parameters. We fit this model to the observed heating rates shown in Fig.\,\ref{fig:hr_vs_location} to find the best-fit values of $\tan\theta$ and $\dot{\overline{n}}_\mathrm{b}$. 

We include an uncertainty estimate for the ion's $z$-location in our heating rate data when fitting Eq.\,\ref{eq:linear_fit}. The ion's position may be shifted along the $z$-axis due to stray electric fields that displace it from the expected trapping location. As an estimate for the stray fields in the $z$-direction, we use the fields along the orthogonal $x$-axis required for micromotion compensation, which amount to $\approx120\,$V/m. From this stray field, we deduce an uncertainty of 7.3$\,$\textmu m for the ion's $z$-location at an axial motional frequency of $\omega=2\pi\times 1\,$MHz. 

\begin{figure}[ht]
    \centering
    \includegraphics[width=3.37in,scale=1.0]{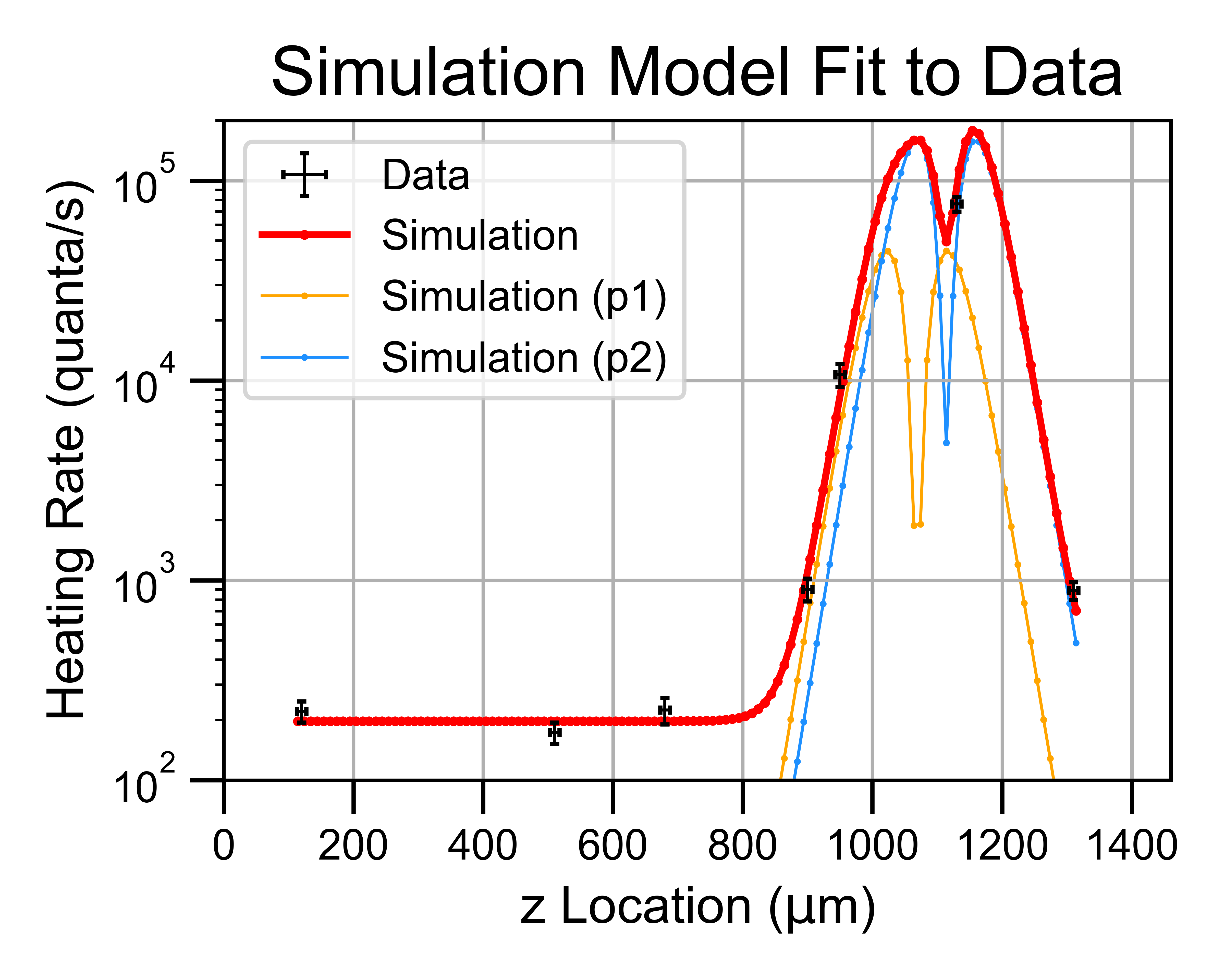}
    \caption{Fit of the lossy dielectric model to motional heating rates measured at various $z$-positions along the trap. Simulation data points are shown as dots, and a cubic spline interpolation is used to generate continuous curves. The orange and blue curves show the noise contributions from p$_1$ and p$_2$, while the black curve shows the combined contribution from p$_1$, p$_2$, and the background heating rate.}
    \label{fig:dielectric}
\end{figure}

The fit produces a reduced chi-squared value  $\chi_\nu^2 = 0.83$ indicating good agreement with the model. Fig.\,\ref{fig:dielectric} shows the model fit to the data, and Table \ref{tab:dielectric} displays the extracted fit parameters. 

\begin{table}[ht]
\caption{\label{tab:dielectric}
Best-fit parameters obtained from fitting Eq.\,\ref{eq:linear_fit} to heating rates measured at different $z$-locations. }
    \begin{ruledtabular}
    \begin{tabular}{c c c}
    Fitting Parameter   &   Value                       \\ [0.5ex] 
    \hline & \\[-2.0ex]
    $\tan\theta$               &   $0.33(0.06)\,$      \\ [0.5ex]
    $\dot{\overline{n}}_\mathrm{b}$               &   $200(10)\,$ ($\mathrm{1/s}$)
    \end{tabular}
    \end{ruledtabular}
\end{table}

The large effective loss tangent of $\tan\theta=0.33(0.06)$ we require to explain our noise magnitude is comparable to loss tangents reported for some thin hydrocarbon polymer films. One study fabricated hydrocarbon films $0.13$–$0.3\,$\textmu m in thickness and reported ($\epsilon_r$, tan$\delta$) = (1.5–3, 0.25) at 1$\,$MHz \cite{nasrin_effect_2020}. Another work\cite{lee_robust_2017} fabricated 30$\,$\textmu m-thick hydrocarbon films and measured ($\epsilon_r$, tan$\delta$) = (1-10, 0.5-2) at 1$\,$MHz. 

\subsection{Analytic Model}
We also compare our results to a simple, analytic model for the particles. The electric field noise produced by a thin dielectric placed on a flat metal surface is given by \cite{kumph_electric-field_2016}:
\begin{eqnarray}
S_\mathrm{E}(\omega) &=&
\bigg(\frac{\tan\theta}{1+\tan^2\theta}\bigg)
\times\bigg(\frac{9 k_\mathrm{B}T}{\pi^2\epsilon_r\epsilon_0\omega}\bigg ) \nonumber\\
&&\times \int_V \frac{(y-y_\mathrm{I})^2(z-z_\mathrm{I})^2}{[(x-x_\mathrm{I})^2+(y-y_\mathrm{I})^2+(z-z_\mathrm{I})^2]^5}dV,
\end{eqnarray}
where the ion's location is ($x_\mathrm{I},y_\mathrm{I},z_\mathrm{I}$) and the integral is performed over volume $V$ of the lossy dielectric.

We consider particle $j$ with location $(x_j,y_j,z_j)$ and dimensions ($\delta x_j,\delta y_j,\delta z_j$) that are small relative to the ion-particle distance, so that the value of the integral may be approximated as:
\begin{eqnarray}
S_{\mathrm{E}}(\omega) &\approx&
\bigg(\frac{\tan\theta}{1+\tan^2\theta}\bigg)\times\bigg(\frac{9 k_\mathrm{B}T}{\pi^2\epsilon_r\epsilon_\omega}\bigg )\nonumber\\
&& \times R(x_j,y_j,z_j)\times(\delta x_j \delta y_j \delta z_j),\nonumber\\
R(x_j,y_j,z_j) &=&
\frac{(y_j-y_\mathrm{I})^2(z_j-z_\mathrm{I})^2}
{[(x_j-x_\mathrm{I})^2+(y_j-y_\mathrm{I})^2+(z_j-z_\mathrm{I})^2]^5}.
\end{eqnarray}

\begin{figure}[ht]
    \centering
    \includegraphics[width=3.37in,scale=1.0]{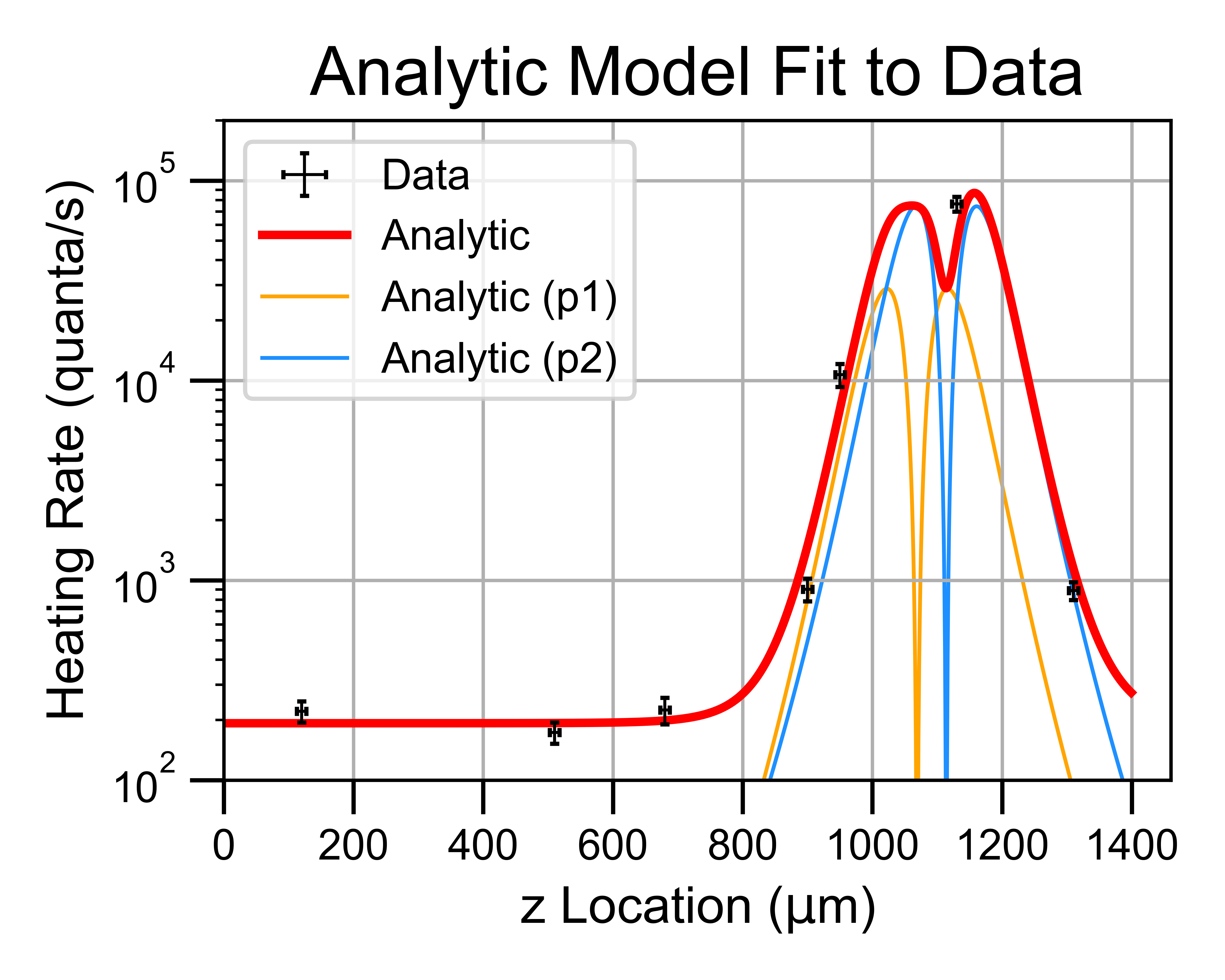}
    \caption{Fit of the analytic model to the motional heating rates measured at various $z$-positions along the trap. }
    \label{fig:analytic_model}
\end{figure}

We fit the analytic model for the particles to our heating rate data. We consider separate contributions $C_{1}R(x_\mathrm{1},y_\mathrm{1},z_\mathrm{1})$ for particle p$_1$ and $C_\mathrm{2}R(x_\mathrm{2},y_\mathrm{2},z_\mathrm{2})$ for particle p$_2$ and include a background heating rate $C_3$:
\begin{equation}\label{eq:nbar}
\dot{\overline{n}} = C_1R(x_{1},y_{1},z_{1})+C_{2}R(x_{2},y_{2},z_{2})+C_3.
\end{equation}
We fit our data to this function by varying the three contribution coefficients $C_i$, again assuming an uncertainty of 7.3$\,$\textmu m for ion $z$-locations. The model fit yields a reduced chi-squared value of $\chi_\nu^2 = 3.4$, indicating that the model is not fully consistent with the data. Nevertheless, the fit qualitatively captures the $z$-dependence of the heating rates. Fig.\,\ref{fig:analytic_model} displays the analytic model fit to the data and the contribution from the two particles. 

\section{Conclusions}
In conclusion, we conjecture that the strong position-dependence of the noise measured above this trap is caused by particles on the surface of the electrodes. The frequency dependence of the noise produced by the particles is approximately $S_\mathrm{E}\propto 1/\omega$. A lossy dielectric model captures the frequency dependence and the position dependence  of the particle-noise. We require a large effective loss tangent of $\tan\theta = 0.33(0.06)$ to explain the magnitude of the noise. 

The observations we make in this publication could explain the wide range of heating rate data reported in literature \cite{brownnutt_ion-trap_2015}. Traps with significantly elevated noise levels may be limited by particles of various sizes, distributions, and materials rather than the intrinsic surface electric field noise that affects all traps. 

\begin{acknowledgments}
We thank the staff of the U.C. Berkeley Marvell Nanolab 
(\url{http://nanolab.berkeley.edu/}) for their help fabricating and imaging the trap.
Support was provided by the
Berkeley NSF-QLCI Challenge Institute for Quantum
Computation (award no. 2016245) and the U.S. Department of Energy, Office of Science, National Quantum Information Science
Research Centers, Quantum Systems Accelerator (QSA, award no. 7562496).
This research used resources of the Advanced Light Source, which is a DOE Office of Science User Facility under contract no. DE-AC02-05CH11231.
\end{acknowledgments}
\section*{Author Declarations}
\subsection*{Conflict of Interest}
The authors have no conflicts to disclose. 

\bibliography{references.bib}

\end{document}